\documentclass[epj]{webofc}
\usepackage[utf8]{inputenc}
\usepackage[varg]{txfonts}   
\usepackage{booktabs}
\usepackage{import}
\usepackage{braket}
\usepackage{multirow}
\usepackage{here}
\usepackage{xcolor}
\definecolor{darkred}{rgb}{0.4,0.0,0.0}
\definecolor{darkgreen}{rgb}{0.0,0.4,0.0}
\definecolor{darkblue}{rgb}{0.0,0.0,0.4}
\usepackage{hyperref}
\usepackage{subfigure}
\wocname{EPJ Web of Conferences}
\woctitle{Lattice2017}
\begin{document}
%
\selectlanguage{english}
\title{%
Vector-Vector Scattering on the Lattice}
\author{%
\firstname{Fernando} \lastname{Romero-López}\inst{1}\fnsep\thanks{Speaker, \email{rlopez@hiskp.uni-bonn.de}} \and
\firstname{Carsten} \lastname{Urbach}\inst{1} \and
\firstname{Akaki}  \lastname{Rusetsky}\inst{1}\fnsep
}
\institute{%
HISKP (Theory) and BCTP, University of Bonn, Germany
}

\abstract{%
 In this work we present an extension of the Lüscher formalism to include the interaction of particles with spin, focusing on the scattering of two vector particles. The derived formalism will be applied to Scalar QED in the Higgs Phase, where the U(1) gauge boson acquires mass.
}
\maketitle
\section{Introduction}\label{intro}

The study of scattering in Lattice Field Theory starts with the original work of Lüscher~\cite{Luscher}. In this first work, he derived equations for the scattering length and phase shift of spinless particles in the rest frame. The formalism has been extended to include moving frames~\cite{Gottt}, $\pi-N$ scattering ~\cite{Bernard:2008ax}, $N-N$ scattering~\cite{Briceno:2013lba}, different masses~\cite{Bernard:2012bi,ZiwenFu}, moving frames with different masses and one particle with spin $1/2$~\cite{Gockeler:2012yj} and any multichannel system with arbitrary spin, momentum and masses~\cite{Briceno:2014oea}.

In this work, we derive a general Lüscher equation for scattering of particles with arbitrary spin through the matching to a non-relativistic effective theory. The results obtained here are in agreement with Ref. \cite{Briceno:2014oea}. We will focus on the case of two identical vector particles and we will make use of the spatial symmetries of the lattice to factorize the Lüscher equation. By means of operators that transform under a certain irreducible representation of the spatial symmetry group, we gain access to the different phase shifts of the theory. The equations will be tested in Scalar QED, for which first numerical results will be shown.

\section{Scattering of two vector particles}\label{sec-1}
\subsection{Derivation of Lüscher Equation for Arbitrary Spin}

Let us consider a system of two particles with masses $M_i$, $i=1,2$ and described by the effective non-relativistic Lagrangian:
\begin{equation}
 \mathcal{L} = \phi_1^\dagger 2 W_1(i\partial_t - W_1)\phi_1 + \phi_2^\dagger 2 W_2(i\partial_t - W_2)\phi_2 + \mathcal{L}_I.
\end{equation}
Here, $\phi_i$ are the non relativistic fields with spin $s_i$ and $W_i = (M_i^2 - \nabla^2)^{1/2}$. The corresponding non-relativistic propagators, with $\omega_i = (M_i^2 + \mathbf{p}^2)^{1/2}$, read
\begin{equation}
 S_i(p) = \frac{1}{2\omega_i(\mathbf{p})} \frac{1}{\omega_i(\mathbf{p})-p^0-i \epsilon},
\end{equation}
The scattering T matrix is defined through the Lippman-Schwinger (LS) equation:
\begin{equation}
 T(z) = (-H_I) + (-H_I)(-G_0(z)) T(z),
 \label{eq:LS}
\end{equation}
where $G_0(z) = (z-H_0)^{-1}$ is the free resolvent and the two particle states coupled to a spin S are
\begin{align}
 &\ket{\mathbf{k_1},\mathbf{k_2},S,\nu} \equiv \ket{\mathbf P,\mathbf k,S,\nu}, \\
 &\braket{\mathbf{P'},\mathbf{k'},S',\nu'|\mathbf P,\mathbf k,S,\nu  } = 4 \omega_1 \omega_2 (2\pi)^d \delta^d(\mathbf{P'}-\mathbf P) (2\pi)^d\delta^d(\mathbf{k'}-\mathbf k) \delta_{S'S} \delta_{\nu'\nu} , \\
 &\mathbf P = \mathbf {k_1}+\mathbf{k_2}, \ \ \ \mathbf k = \mu_2 \mathbf{k_1} - \mu_1 \mathbf{k_2}, \ \ \ \mu_{1,2} = \frac{1}{2}\left(1\pm \frac{m_1^2- m_2^2}{ P^2}\right),
 \label{eq:statesLab}
\end{align}
where $S$, $\nu$ label the total spin of the two particle system and $\mathbf P$, $\mathbf k$ are the total and relative momentum in the ``laboratory frame''. Now define the matrix elements:
\begin{equation}
 t^{S'S}_{\nu'\nu}(\mathbf{k'},\mathbf k,\mathbf P,z) =\int \frac{d^d\mathbf {P'}}{(2\pi)^d} \braket{\mathbf{P'},\mathbf{k'},S',\nu'|T(z)|\mathbf P,\mathbf k,S,\nu}, \label{eq:tnunu}
\end{equation}
\begin{equation}
 h^{S'S}_{\nu'\nu}(\mathbf{k'},\mathbf k,\mathbf P) = \int \frac{d^d\mathbf{P'}}{(2\pi)^d} \braket{\mathbf{P'},\mathbf{k'},S',\nu'|(-H_I)|\mathbf P,\mathbf k,S,\nu}. \label{eq:hnunu}
\end{equation}
Additionally, $G_0$ can be written in terms of an elementary diagram using the Feynman rules and
one can perform the integration over $q^0$ using the Cauchy integration formula. One may rewrite the Lippman-Schwinger Equation in terms of matrix elements using Equations \ref{eq:tnunu} and \ref{eq:hnunu}:
\begin{align}
 t^{S'S}_{\nu'\nu}(\mathbf{k'},\mathbf k,\mathbf P,z) = h^{S'S}_{\nu'\nu}(\mathbf {k'},\mathbf k,\mathbf P) + \int \frac{d^d\mathbf{q_1}}{(2\pi)^d} \sum_{\nu''} \frac{  h^{S'S''}_{\nu'\nu''}(\mathbf{k'},\mathbf{q_1},\mathbf P) t^{S''S}_{\nu''\nu}(\mathbf{q_1},\mathbf k,\mathbf P,z)}{4\omega_1(\mathbf{q_1})\omega_2(\mathbf P-\mathbf{q_1}). (\omega_1(\mathbf{q_1})+\omega_2(\mathbf{P}-\mathbf{q_1})-z)}, \label{eq:LS2}
\end{align}
where $\mathbf q = \mu_2 \mathbf{q_1} - \mu_1 \mathbf{q_2} $, as in Equation \ref{eq:statesLab}. A key point here is that the elementary bubble is diagonal in spin, because also the single particle propagators are. However, the scattering amplitude need not be diagonal.
Now define the projectors to the partial waves in the CM frame, whose momenta are $\mathbf{k^*}$:
\begin{align}
 &\Pi_{\nu'\nu}^{A'A} ( \mathbf {k'^*}, \mathbf{ k^*}) =\sum_{\rho, \rho'} U^{S}_{\nu'\rho'}(\mathbf{k'^*})^* U^{S}_{\nu \rho}(\mathbf{k^*}) (\mathcal{Y}_{J'l'S'\mu'}(\mathbf{k'^*},\rho'))^* \mathcal{Y}_{JlS\mu}(\mathbf{ k^*},\rho), \\
 &A=(J,l,S,\mu), \ \ A'=(J',l',S',\mu'), \label{eq:multiindex}
\end{align}
where $U^{S}_{\nu \rho}(\mathbf{k^*})$ is the unitary transformation of the spin indices under a boost and the spin spherical harmonics, with $\mathbf{\hat k}= \mathbf k / |\mathbf{k}|$, read
\begin{equation}
 \mathcal{Y}_{JlS\mu}(\mathbf k,\nu) = \sum_{m,\sigma} \braket{l m S \sigma | J\mu} |\mathbf k|^l Y_{lm}(\mathbf{\hat k}) \chi^S_{\sigma}(\nu) \equiv |\mathbf k|^l Y_{JlS\mu}(\mathbf{\hat k},\nu). \label{eq:sphar}
\end{equation}
Using the projectors, Equations \ref{eq:tnunu} and \ref{eq:hnunu} take the form
\begin{align}
t^{S'S}_{\nu'\nu}(\mathbf{k'},\mathbf k,\mathbf P,z) &= 4\pi \sum_{A',A} \Pi_{\nu'\nu}^{A'A}(\mathbf{k'^*},\mathbf{k^*}) t_{A'A}(s,\mathbf P,z),\label{eq:texpansion} \\
h^{S'S}_{\nu'\nu}(\mathbf{k'},\mathbf{k},\mathbf P) &= 4\pi \sum_{A',A} \Pi_{\nu'\nu}^{A'A}(\mathbf{k'^*},\mathbf{k^*}) h_{A'A}(s, \mathbf P).\label{eq:hexpansion}
\end{align}
If the system is placed in a box, the integral may be replaced by an sum:
\begin{equation}
 \int \frac{d^d\mathbf {q_1}}{(2\pi)^d}  \rightarrow \frac{1}{L^3}\sum_{\mathbf{q_1}},
 \label{eq:sumint}
\end{equation}
and by plugging Equations \ref{eq:texpansion} and \ref{eq:hexpansion} into Equation \ref{eq:LS2} on the mass-shell, one arrives to
\begin{align}
 t_{A'A}(s,\mathbf P) - h_{A'A}(s,\mathbf P) = \frac{k^*}{8\pi \sqrt{s}}\sum_{B,B'} h_{A'B}(s,\mathbf P)( (\mathbf{k^*})^{l+l'} i^{l-l'} \delta_{S_B S_{B'}} \mathcal{M}_{BB'}(s,\mathbf P)) t_{B'A}(s,\mathbf P), 
 \label{eq:LS3}
\end{align}
with $s=P^2$ and $S_B$ being the spin of the multi-index $B$ as in Equation \ref{eq:multiindex}. Now, using unitarity of the transformation of the spin indices, one arrives at
\begin{align}
 \mathcal{M}_{JlS\mu,J'l'S'\mu'}(s,\mathbf P) = \frac{32\pi^2}{\mathbf{k^*}} \frac{\sqrt{s}}{L^3} i^{l'-l} \delta_{S'S} \sum_{q_1} \frac{\sum_\nu({Y}_{JlS\mu}(\mathbf{\hat q^*},\nu))^*{Y}_{J'l'S\mu'}(\mathbf{\hat q^*},\nu)}{4\omega_1(\mathbf{q_1})\omega_2(\mathbf P-\mathbf{q_1}) (\omega_1(\mathbf{q_1})+\omega_2(\mathbf P-\mathbf{q_1})-P_0)}.
\end{align}
This matrix can be related to its equivalent for scalar particles by using Equation \ref{eq:sphar}:
\begin{equation}
 \mathcal{M}_{JlS\mu,J'l'S'\mu'} = \delta_{S'S} \sum_{m,m',\sigma} \braket{ lm,S\sigma | J \mu} \braket{ l'm',S\sigma | J' \mu'} \mathcal{M}_{lm,l'm'},
\end{equation}
where we used the identity \cite{Gasser:2011ju,Bernard:2012bi} (with $\mathbf q = \mathbf{q_1} - \mu_1 \mathbf P$)
\begin{align}
\begin{split} 
& \frac{1}{4\omega_1\omega_2 (\omega_1+\omega_2-P_0)} = \frac{1}{2P_0} \frac{1}{\mathbf q^2-\frac{(\mathbf q \mathbf P)^2}{P_0^2}-(\mathbf{k^*})^2}  \\
&+ \frac{1}{4\omega_1\omega_2}\left( \frac{1}{\omega_1+\omega_2+P_0} - \frac{1}{\omega_1-\omega_2+P_0} - \frac{1}{\omega_2-\omega_1+P_0}   \right),
\label{eq:iden}
\end{split}
\end{align}
kept only the divergent part (first term in Equation \ref{eq:iden}) and used $(\mathbf{q^*})^2= \mathbf{q}^2 - \frac{(\mathbf{qP})^2}{P^2_0}$.
This way, and up to exponentially suppressed terms ($\mathbf{q^*} \rightarrow \mathbf{k^*}$), $\mathcal{M}_{lm,l'm'}$ is given by (see \cite{Bernard:2012bi})
\begin{equation}
 \mathcal{M}_{lm,l'm'}(\mathbf{k^*},s) = \frac{(-1)^l}{\pi^{3/2}\gamma} \sum_{j=|l-l'|}^{l+l'}\sum_{s=-j}^j \frac{i^j}{\eta^{j+1}} Z^d_{js}(1,s)^* C_{lm,js,l'm'},  \ \eta= \frac{|\mathbf{k^*}|L}{2\pi}, 
\end{equation}
where
\begin{equation}
 C_{lm,js,l'm'} = (-1)^{m'} i^{l-j+l'} \sqrt{(2l+1)(2l'+1)(2j+1)}
 \begin{pmatrix}
 l & j & l' \\ m & s & -m'
 \end{pmatrix}
 \begin{pmatrix}
 l & j & l' \\ 0 & 0 & 0
 \end{pmatrix},
\end{equation}
\begin{equation}
 Z^d_{lm}(1,s) = \sum_{\mathbf r\in P_d} \frac{|\mathbf r|^l Y_{lm}(r)}{\mathbf r^2-\eta^2}, \ \ \ P_d=\{\mathbf{r_{||}}= \gamma^{-1}(\mathbf{n_{||}}-\mu_1 \mathbf d), \mathbf{r_{\perp}}= \mathbf{n_{\perp}}   \},
 \label{eq:defZ}
\end{equation}
with $n \in \mathbb{Z}$. One can see that Equation \ref{eq:LS3} is a matrix equation, and the poles in $t_{A'A}$ arise when
\begin{equation}
 \det  \left( \frac{8\pi \sqrt{s}}{(\mathbf{k^*})^{l+l'+1}} (h^J_{lS,l'S'})^{-1} \delta_{JJ'} \delta_{\mu\mu'}-  \delta_{S,S'} \mathcal{M}_{JlS\mu,J'l'S\mu'}\right) =0,
 \label{eq:luscher1}
\end{equation}
where it is already implied that $J$ and $\mu$ are conserved in scattering processes in infinite volume and mixings can be present in $l$ and $S$, $h_{JlS\mu,J'l'S'\mu'} = h^J_{lS,l'S'}\delta_{JJ'}\delta_{\mu\mu'}$. In order to express this equation in a more compact way, one uses the standard definition of the $S$ matrix (see \cite{Briceno:2013bda}, for nucleon-nucleon scattering), $S = e^{2i\delta(s)}$, in terms of the phase shift $\delta(s)$. This way, one can express $h^J_{lS,l'S'}$ in terms of $\delta$:
\begin{equation}
 h^J_{lS,l'S} = \frac{8\pi\sqrt{s}}{(\mathbf{k^*})^{l+l'+1}} (\tan \delta)^J_{lS,l'S'}, \label{eq:hAA}
\end{equation}
and plugging it in Equation \ref{eq:luscher1}, we arrive at
\begin{equation}
 \det  \left( (\cot \delta)^J_{lS,l'S'} \delta_{JJ'}\delta_{\mu\mu'}-\delta_{SS'} \mathcal{M}_{JlS\mu,J'l'S\mu'}\right) =0.
 \label{eq:luscher2}
\end{equation}

\subsection{Two Vector Particles \label{sec:twovector}}

A system of two identical vector particles can couple to total spin $S=0,1,2$. Even spin combinations are symmetric under the exchange of two particles, whereas odd combinations are antisymmetric. The same holds for the angular momentum $L$. The possible combinations of $S$ and $L$ to $J^P$ respecting the Bose statistics (totally symmetric state) are listed in the Table \ref{tab:Jp}. The combinations that have mixing are in the same column in the table and correspond to same $J^P$ but different $L$, $S$.
\begin{table}[H]
\centering
\begin{tabular}{| c | c | c | c | c | c | c | }
\hline
 $J^P$                     &$0^+$    & $0^-$   & $1^+$   & $1^-$   & $2^+$   & $2^-$\\ \hline
 \multirow{3}{*}{$\{S,L\}$}&$\{0,0\}$&         &         &         &$\{0,2\}$& \\ \cline{2-7}
                           &         &$\{1,1\}$&         &$\{1,1\}$&         &$\{1,1\}$,$\{1,3\}$ \\ \cline{2-7}
                           &$\{2,2\}$&         &$\{2,2\}$&         &$\{2,0\}$,$\{2,2\}$,$\{2,4\}$&\\ \cline{1-7}
\end{tabular}  
\caption{Possible values of $J^P$ with $J<3$. \label{tab:Jp}}
\end{table}
The possible mixings can be parametrized with a mixing angle and two eigenvalues. This would be analogous to the parametrization of the mixings for two nucleons in Reference \cite{Briceno:2013bda}.




For the scattering of two spinless particles, it is well known that the phase shift can be parametrized as a polynomial of $\mathbf k^2$.
Such an equivalent parametrization can be derived here as well:
\begin{equation}
 \mathbf k^{l+l'+1} \cot \delta^J_{lS,l'S'} = \sum_{n=0} a_n \mathbf k^{2n}.
\end{equation}


\subsection{Reduction of Lüscher Equation}
The basis vector labeled by $\alpha$ of a symmetry group $\mathcal{G}$, in a certain irreducible representation $\Gamma$, are constructed (up to normalization) applying the projector
\begin{equation}
( P^{\Gamma,J,l}_{\alpha\beta})_{\mu\mu'} = \sum_{i \in G} (R_i^\Gamma)^*_{\alpha\beta} \tilde{ D}^{J,l}_{\mu\mu'}(R_i),
\end{equation}
to the basis vectors in the continuum, $\ket{J,S,l,\mu}$, for a fixed $\beta$ and $\mu$:
\begin{equation}
 \ket{\Gamma, \alpha, J, S, l, n} \propto \sum_{\mu'} ( P^{\Gamma,J,l}_{\alpha\beta})_{\mu\mu'} \ket{J,S,l,\mu'},
\end{equation}
where $n$ labels the number of occurrences of $\Gamma$ in $J$, $l$. In the previous equations $\tilde{ D}^{J,l}_{\mu\mu'}(R_i) = (-1)^{l} { D}^J_{\mu\mu'}(R_i)$ if the element $i$ includes inversion, or just the standard Wigner matrix if not. The basis vectors of the irreducible representations of the symmetry group of the lattice can be expressed in terms of the one of the continuum:
\begin{equation}
 \ket{\Gamma,\alpha,J,l,S,n} = \sum_{\mu} c^{\Gamma n \alpha}_{Jl\mu} \ket{JlS\mu},
\end{equation}
where the coefficients $c^{\Gamma n \alpha}_{Jl\mu}$ are to be read from basis vector tables in References \cite{Gockeler:2012yj} and \cite{Bernard:2008ax}, for example. The matrix $\mathcal{M}$ can be partially diagonalized in this basis:
\begin{align}
  &\braket{\Gamma',\alpha',J',l',S,n'| \mathcal{M}  |\Gamma,\alpha,J,l,S,n} = \mathcal{M}^{\Gamma}_{J'l'Sn',JlSn} \delta_{\Gamma'\Gamma} \delta_{\alpha' \alpha},  \\
  &\mathcal{M}^{\Gamma}_{J'l'Sn',JlSn} = \sum_{\mu'\mu} ( c^{\Gamma n' \alpha}_{J'l'\mu'})^*  c^{\Gamma n \alpha}_{Jl\mu} \mathcal{M}_{J'l'S\mu',JlS\mu}.
\end{align}
Moreover, the matrix $\cot \delta$ has to be brought to the same basis as $\mathcal{M}$:
\begin{equation}
(\cot \delta)^{\Gamma}_{JlSn,J'l'S'n'}=\sum_{\mu'\mu} ( c^{\Gamma n' \alpha}_{J'l'\mu'})^*c^{\Gamma n \alpha}_{Jl\mu} (\cot \delta)^J_{lS,l'S'}\delta_{\mu\mu'}\delta_{JJ'}  =\sum_{\mu} ( c^{\Gamma n' \alpha}_{Jl'\mu})^*  c^{\Gamma n \alpha}_{Jl\mu} (\cot \delta)^J_{lS,l'S'} \delta_{JJ'}.
\end{equation}
The coefficients $c^{\Gamma n \alpha}_{Jl\mu}$ just depend on the total angular momentum $J$ and the behaviour under spatial inversions $(-1)^l$. Moreover, mixing can only occur between states with same $J$ and $(-1)^l$, so the sum over the index $\mu$ yields either 1 or the matrix $\cot \delta$ is trivially zero due to the symmetry considerations. Spin does not enter here at all, because it does not influence the coefficients. This way, $(\cot \delta)^{\Gamma}_{JlSn,J'l'S'n'}$ is diagonal in $J$ and $n$
and the determinant factorizes:
\begin{equation}
\prod_{S= \substack{even\\odd}}  \ \  \prod_{\Gamma} \ \det  \left( (\cot \delta)^J_{lS,l'S'}\delta_{JJ'}\delta_{nn'}-\delta_{SS'}\mathcal{M}^{\Gamma}_{JlSn,J'l'Sn'}\right) =0.
\end{equation}

\section{Toy Model: Scalar QED \label{sec:toymodel}}

In order to test the formalism, we use Scalar QED with a Higgs mechanism, since the vector state needs to be massive. The continuum Euclidean Lagrangian of such a theory reads
\begin{equation}
 \mathcal{L}_E = (D_\mu \phi_c)^\dagger D_\mu \phi_c + m^2_0 |\phi_c|^2 +\lambda_c |\phi_c|^4, \quad D_\mu \phi_c = \partial_\mu \phi_c +i g A_\mu \phi_c. \label{eq:toymodel}
\end{equation}
The discretized action is (See \cite{SCALARQED}) 
\begin{align}
 S = \sum_x &\Big(-\frac{\beta}{2}\sum_{\mu<\nu}(U_{\mu\nu} + U_{\mu\nu}^*) -\kappa \sum_\mu(\phi_x^* U_{x,\mu}\phi_{x+\mu} +cc) + \lambda(|\phi_x|^2 - 1)^2 +  | \phi_x|^2  \Big),
\end{align}
with
\begin{align}
 \lambda_c = \frac{\lambda}{\kappa^2}, \quad (a m_{0})^2 =   \frac{1-2\lambda-8\kappa}{\kappa}, \quad \beta = \frac{1}{g^2}.
\end{align}
The basic operators for the scalar and real particle are
\begin{align}
 &\mathcal{O}^{A_1} =\sum_i \operatorname{Re}(\phi^\dagger(x,t) U_{i}(x) \phi(x + ae_i,t)), \label{eq:0+1} \\
 &\mathcal{O}^{T_1^{-}}_{n,i} = \operatorname{Im}(\phi^\dagger(x,t) \left( \prod_{n=0}^{N-1} U_{i}(x+ne_i) \right) \phi(x + Ne_i,t)), \label{eq:1-1}
\end{align}
respectively. Notice that the second one is highly non-local, since this have been seen to improve significantly the signal. In particular, with the vector operator, one can build one and two particle operators that transform under certain irreducible representations of the spatial symmetry group, in rest and moving frames. A one particle operator transforming under a certain irrep $\Gamma$ with invariant momentum under the spacial symmetry group $p \in \mathcal{G}$ can be built from an arbitrary operator $\mathcal{O}(x,t)$ as follows:
\begin{equation}
 \mathcal{O}^\Gamma (p, t) = \sum_{x} e^{ipx} \sum_{S_i \in \mathcal{G}} \chi^\Gamma(S_i) \ (S_i \mathcal{O})(S_i^{-1}x ,t )
\end{equation}
Equivalently, for a two particle operator:
\begin{equation}
 \mathcal{O}^\Gamma (p, q, t) = \sum_{x,y} \left( \sum_{S_i \in \mathcal{G}} e^{ipx + S_iq(y-x)} \right)  \sum_{S_i \in \mathcal{G}} \chi^\Gamma(S_i) \ (S_i \mathcal{O}) (S_i^{-1}x,S_i^{-1}y ,t ), \label{eq:op2part}
\end{equation}
where $p$ is the total momentum that belonging to the spatial symmetry group $\mathcal{G}$, $q$ is a relative momentum that the two particles have and $\mathcal{O}(x,y,t)$ is an arbitrary two particle operator.

\section{Numerical Results}
\begin{figure}[tp]
   \centering
   \subfigure[Mass of the vector particle for ensemble A12 for different lengths of the operator in Equation \ref{eq:1-1}. \label{fig:imL}]%
             {\includegraphics[width=0.475\textwidth,clip]{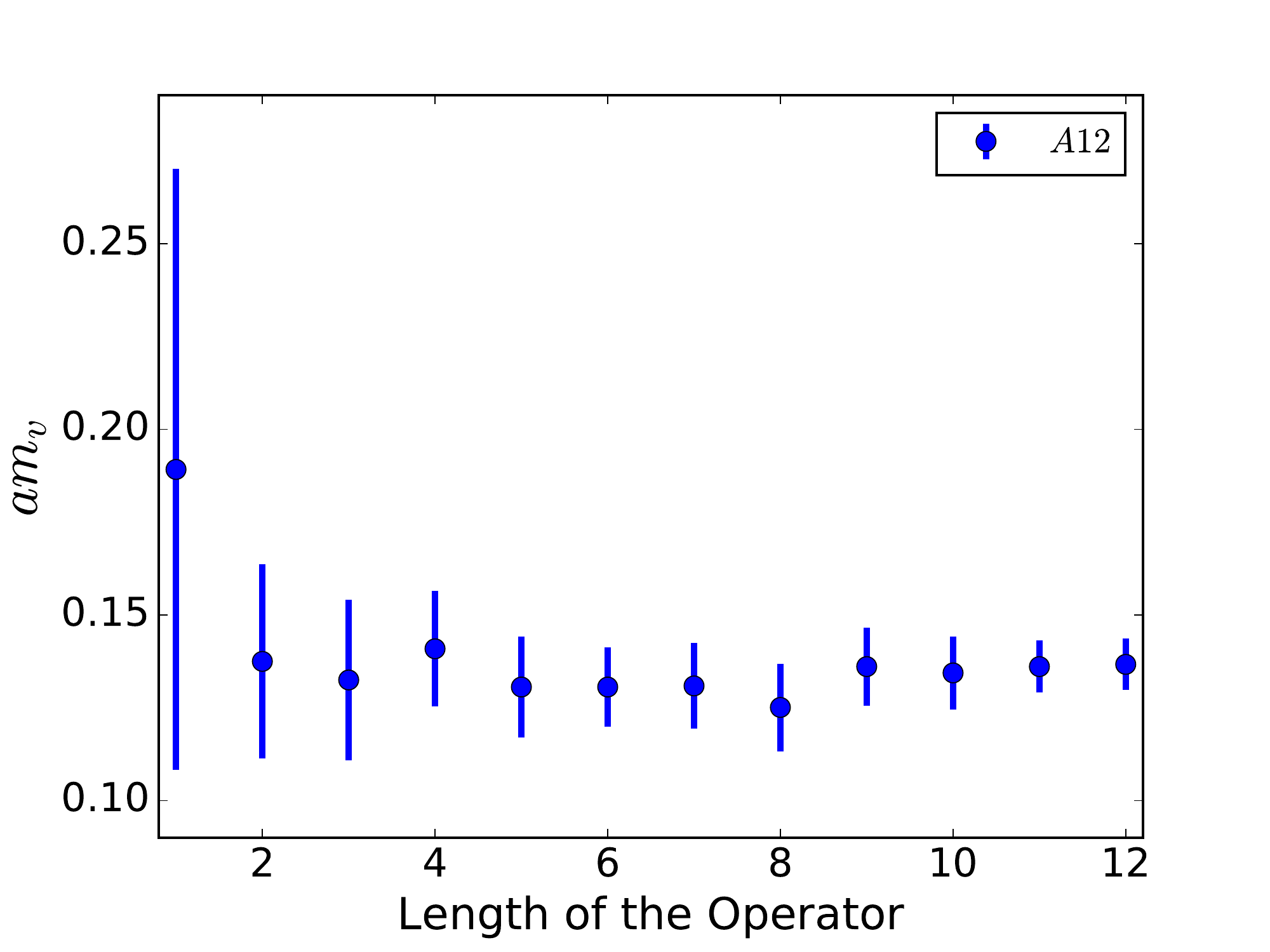}}\hfill
   \subfigure[Mass of the scalar and vector particle for $L=12$ as a
     function of $\kappa$. \label{fig:spectrum}]%
             {\includegraphics[width=0.475\textwidth,clip]{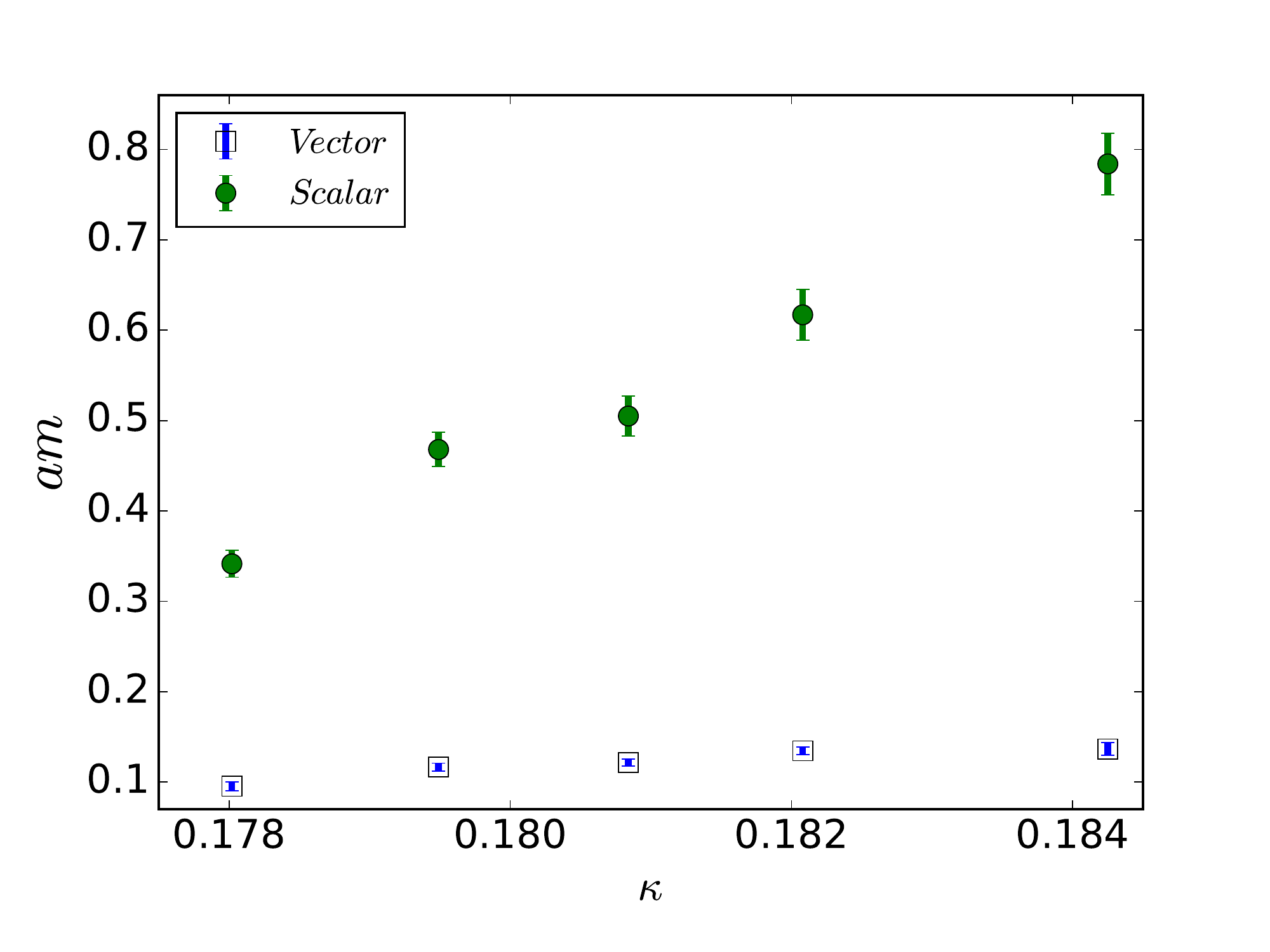}}
   \subfigure[Energy difference $\Delta E$ as a function of $\kappa$ in the rest frame.  \label{fig:diffE}]%
             {\includegraphics[width=0.475\textwidth,clip]{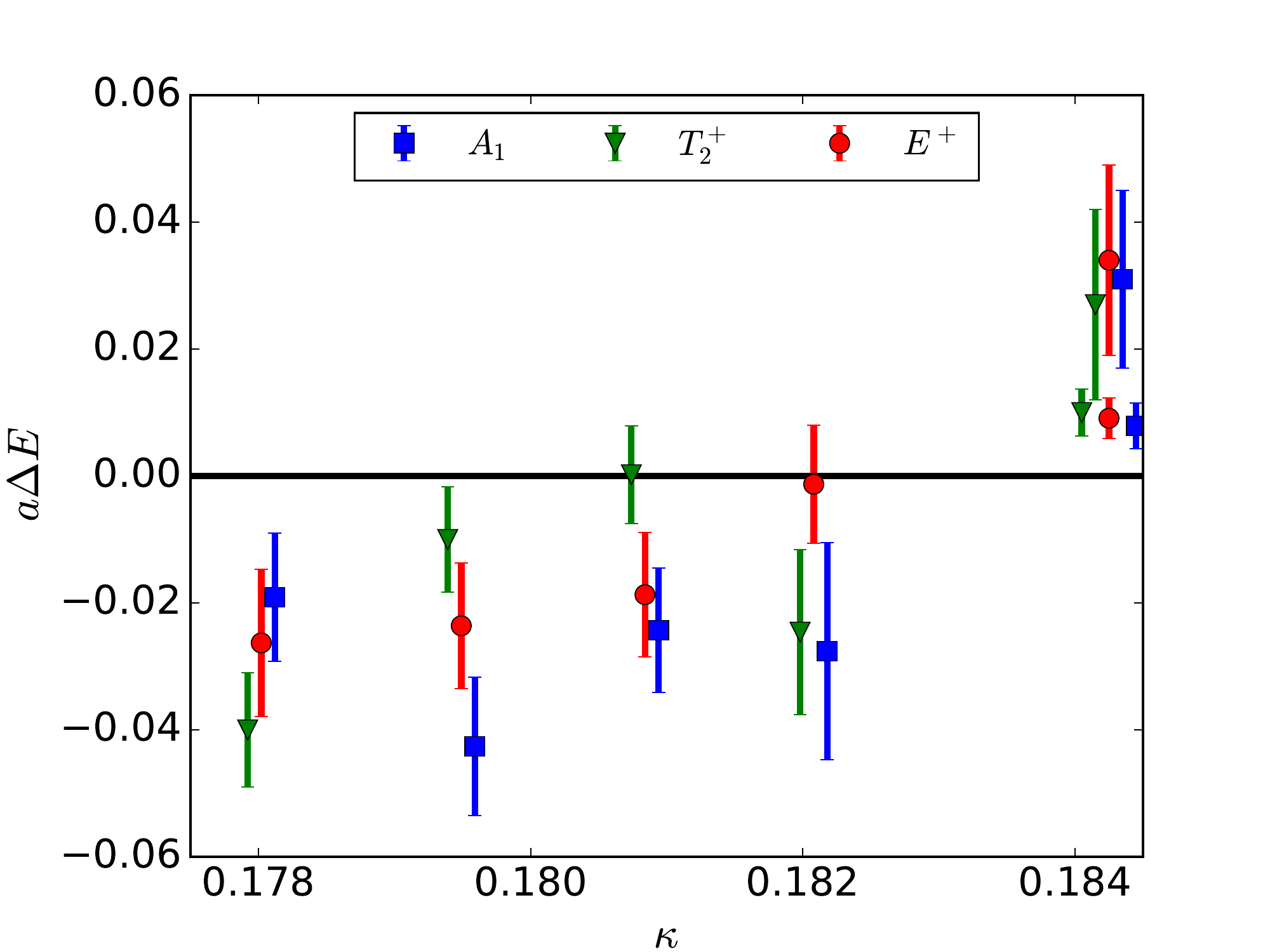}} \hfill
   \subfigure[Mass of the vector particle for different irreps for ensemble A16. \label{fig:moving16}]%
             {\includegraphics[width=0.475\textwidth,clip]{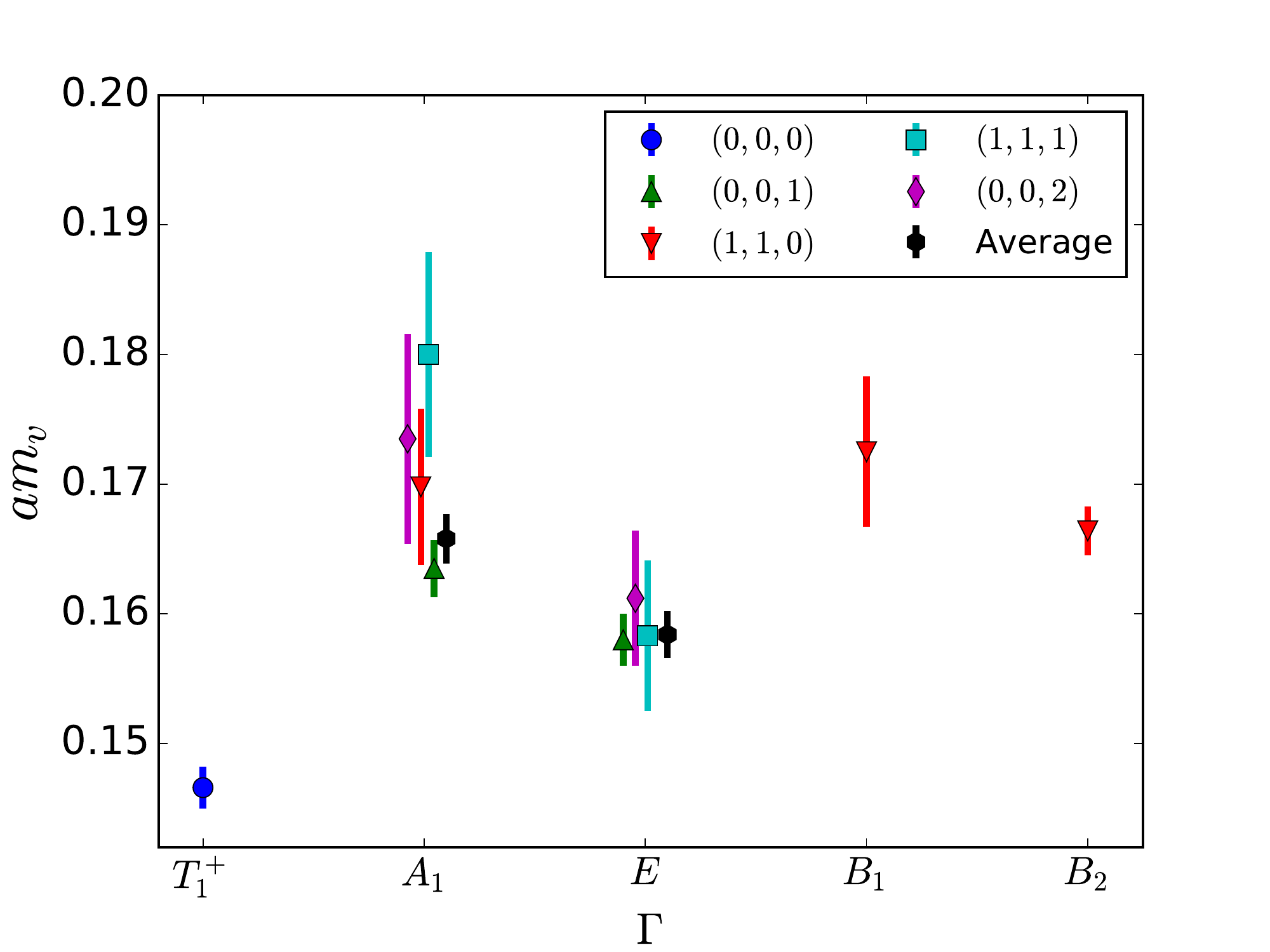}}
   \subfigure[Energy difference $\Delta E$ for ensemble A16 in the
     moving frame with total momentum $p=\frac{2\pi}{L}(0,0,1)$ and
     $q=0$ as in Equation \ref{eq:op2part} for different irreps $\Gamma$. \label{fig:001}]%
             {\includegraphics[width=0.475\textwidth,clip]{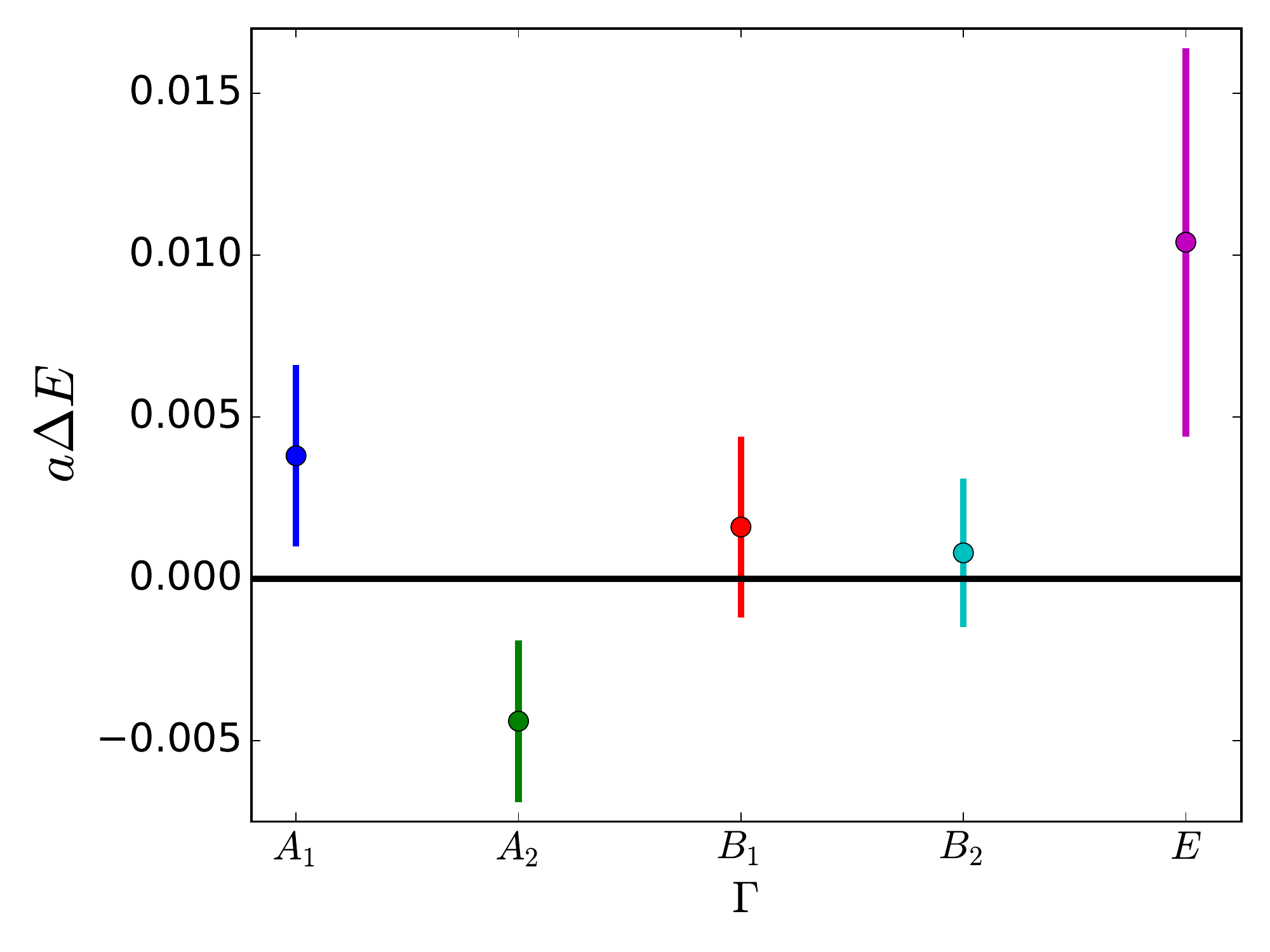}}\hfill
   \subfigure[Phase shift in the $J^P=0^+$ channel as a function of
     the scattering momentum $k$. They are calculated neglecting partial waves $J>1$ and at this level, the two possible $L,S$ combinations cannot be distinguished. \label{fig:phaseshift}]%
             {\includegraphics[width=0.475\textwidth,clip]{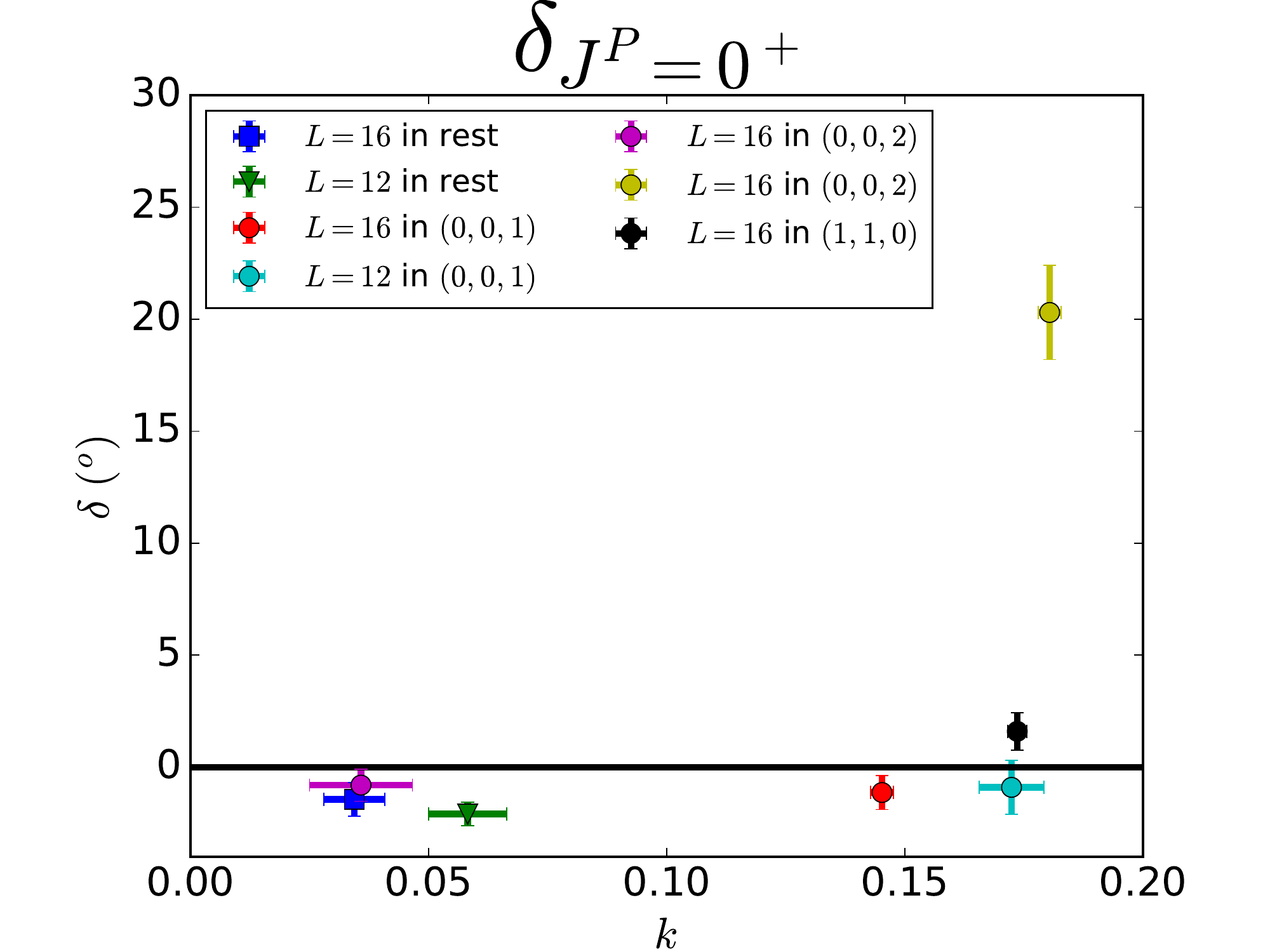}}
   \caption{Numerical results.}
   \label{fig:LICup}
\end{figure}

We use five different parameters for $L=12$ and for one parameter set
$L=12$ and $L=16$ (see Table \ref{tab:ensembles12}). In
Figure~\ref{fig:imL} we show the dependence of mass of the vector
particle on the length of the operator in Equation~\ref{eq:1-1} for
ensemble A12. The results show a clear 
improvement of the signal with non-local operators. When using moving
frames, the best signal is empirically seen at $N = L/(d+1)$, being
$d$ the units of momentum in that particular direction.  

Moreover, we study the mass of a single vector and scalar particle in
the $L=12$ volume for the different bare parameter sets (Figure
\ref{fig:spectrum}). In the continuum, the bare masses of the
particles are $m^2_\phi = -2m_0^2$ and $m_V^2 =
-\frac{g^2m_0^2}{\lambda_c}$. The mass of the vector is suppressed by
$g$ and $\lambda_c$, and it seems logical that it is much smaller than
the scalar mass. However, it is not clear why the scalar mass
duplicates with increasing $\kappa$, whereas the vector mass increases
slightly. Naïvely, the increase of the vector mass should be enhanced
by the reduction of $\lambda_c$ with increasing $\kappa$, but there
seems to be some non-trivial behaviour.  

The energy difference $\Delta E$ is defined as the difference between
the two particle energy on the lattice and the two particle energy in
absence of interactions. In Figure \ref{fig:diffE} we show the results
for $\Delta E$ as a function of $\kappa$. For higher $\kappa$, the
interaction generates scattering states ($\Delta E >0$) with
$\gtrapprox 2\sigma$ statistical significance. As $\kappa$ becomes
smaller, all particles become lighter and the interaction seems to
flip sign. For the lowest values of $\kappa$, the two particle states
are bound states ($\Delta E <0$), with a similar statistical
significance as in the other case. Unfortunately, in the transition
region the statistical significance is rather small. Besides, we see
the expected volume dependence in the energy shift, when comparing
$L=12$ with $L=16$. 

In addition, we show the results of the mass of a single vector
particle for the different irreps in Figure \ref{fig:moving16}. The
values for the different moving frames tend to always larger than the
one in the rest frame. Since the correlation functions are fitted to
include an excited state, a contamination of higher states is likely
not the main reason. It is however true that from the correlation
functions one obtains upper bounds for the energy, which might have an
influence. In addition, the different irreps seem to split. In
particular $A_1$ and $E$ seem to differ by around $3\sigma$. 

Furthermore, in Figure \ref{fig:001} we show $\Delta E$ for the
ensemble A16 in the first moving frame. One can see how the different
irreps split. With these values one can calculate phase shifts values
(Table \ref{tab:phase001}). Finally, in Figure \ref{fig:phaseshift} we
show all our results for the phase shift with $J^P=0^+$. For the
highest momentum 
shown, the ratio between the non-relativistic kinetic energy and the
mass is quite large, $\frac{E_k}{m} \approx 0.8$. Hence, the kinematic
suppression of higher partial waves, though present, is not strong any
more and a corresponding systematical error is to be expected. Still,
it seems that the phase shift increases for the last points. If this
increase is physical, it could correspond to a resonance around
scattering momentum $ak \approx 0.18$. 

\begin{center}
\begin{table}[H]
\centering
\begin{tabular}{| c | c | c | c | c | c | c | }
\hline
$L^3 \times T$  & ref. & N    &$m^2_0$ & $\lambda_c$ & $\kappa$ & $\lambda$ \\ \hline
$16^3 \times 24$&   A16   & 138000 &-35     & 88          & 0.18425 & 2.9873    \\ \cline{1-7}
\multirow{5}{*}{$12^3 \times 24$}&   A12   & 33000 &-35     & 88          & 0.18425 & 2.9873    \\ \cline{2-7}
&   B12   & 24000 &-35.5     & 90.6          & 0.18208 & 3.0036    \\ \cline{2-7}
&   C12   & 21500 &-35.6     & 91.6          & 0.18084 & 2.9956    \\ \cline{2-7}
&   D12   & 16000 &-35.85    & 93.1          & 0.17949 & 2.9994    \\ \cline{2-7}
&   E12   & 18800 &-36.1     & 94.7          & 0.17802 & 3.0012    \\ \hline
\end{tabular}  
\caption{Ensembles used for the simulations. The gauge coupling is kept constant, $\beta = 2.5$. \label{tab:ensembles12}}
\end{table}
\end{center}
\begin{minipage}[b]{1\linewidth} 
\centering
\begin{table}[H]
\centering
\begin{tabular}{| c | c | c | c |}
\hline
 $\Gamma$ & $ak$ & $J^P$ & $\delta \ (^o)$      \\   \hline
 $A_1$   &0.1452(24)& $0^+$  &-1.13(76) \\ \hline 
 $A_1$   &0.1452(24)& $1^-$ &-0.14(10) \\ \hline 
 $A_2$   &0.1369(21)& $0^-$ &1.24(67)  \\ \hline 
 $A_2$   &0.1369(21)& $1^+$ &0.127(60)  \\ \hline 
\end{tabular}  
\caption{Obtained values for the phase shift in moving frame with the assumption of no mixings. The results are in agreement with the kinematic suppression expected for higher partial waves. \label{tab:phase001}}
\end{table}
\end{minipage}

\section{Summary and Outlook}

We have shown that the study on the Lattice of scattering of vector particles is possible by applying the derived framework to the toy model Scalar QED. The complete numerical results, together with the group theoretical derivations, will be published soon in a longer, more detailed version. On the long term, we expect to apply the ideas of this work to study the possibility of the Higgs boson to be a resonance of two $W$ bosons. This is the case for a model proposed by Frezzotti \textit{et al.}~ \cite{Frezzotti:2014wja,Frezzotti:2016bes}, where a ``superstrong interaction'' together with superstrongly interacting particles are present.

We would like to acknowledge the lattice group in Bonn and Roberto
Frezzotti for the interesting discussions and the support
provided. This work was supported in part by the DFG in the
Sino-German CRC110. Finally, special thanks to BCGS for the continuous support.

\bibliography{lattice2017}

\end{document}